\documentclass[11pt,letterpaper]{article}

\usepackage{times}
\usepackage{helvet}
\usepackage{courier}
\usepackage[T1]{fontenc}

\usepackage[margin=1in]{geometry}

\usepackage{amsmath}
\usepackage{amsfonts}
\usepackage{amssymb}

\usepackage{booktabs}
\usepackage{graphicx}
\usepackage{caption}

\usepackage[hyphens]{url}
\usepackage[hidelinks,colorlinks=true,linkcolor=blue,citecolor=blue,urlcolor=blue]{hyperref}
\usepackage{natbib}
\usepackage[dvipsnames]{xcolor}
\usepackage{comment}
\usepackage{siunitx}
\usepackage{enumitem}
\usepackage{float}
\usepackage{algorithm}
\usepackage{algorithmic}
\usepackage{newfloat}
\usepackage{listings}
\usepackage{bibentry}

\floatstyle{ruled}
\newfloat{listing}{tb}{lst}{}
\floatname{listing}{Listing}

\title{SELDON: Supernova Explosions Learned by Deep ODE Networks}

\author{
Jiezhong Wu$^{1,2,*}$,\quad
Jack O'Brien$^{1,3,*,\dagger}$,\quad
Jennifer Li$^{1,3,4}$,\quad
M.~S.~Krafczyk$^{1,4}$,\\[4pt]
Ved G.~Shah$^{1,5,6}$,\quad
Amanda R.~Wasserman$^{1,3,4}$,\quad
Daniel W.~Apley$^{1,2}$,\\[4pt]
Gautham Narayan$^{1,3,4,\ddagger,\dagger}$,\quad
Noelle I.~Samia$^{1,7,\ddagger,\dagger}$
\\[10pt]
\normalsize $^{1}$NSF-Simons AI Institute for the Sky (SkAI), Chicago, IL, USA\\
\normalsize $^{2}$Department of Industrial Engineering and Management Sciences, Northwestern University\\
\normalsize $^{3}$Department of Astronomy, University of Illinois Urbana-Champaign\\
\normalsize $^{4}$National Center for Supercomputing Applications (NCSA), Urbana, IL, USA\\
\normalsize $^{5}$Department of Physics and Astronomy, Northwestern University\\
\normalsize $^{6}$Center for Interdisciplinary Exploration and Research in Astrophysics (CIERA), Northwestern University\\
\normalsize $^{7}$Department of Statistics and Data Science, Northwestern University
\\[8pt]
\normalsize $^{*}$Equal contribution.\quad
$^{\ddagger}$Equal contribution.\quad
$^{\dagger}$Corresponding authors.
\\[4pt]
\small\texttt{\{jiezhongwu2021, vedshah2029, apley\}@northwestern.edu}\\
\small\texttt{\{jackob, jli184, mkrafcz2, amandaw8, gsn\}@illinois.edu}\\
\small\texttt{n-samia@northwestern.edu}
}

\date{}

\begin{document}
\maketitle

\begin{abstract}
The discovery rate of optical transients will explode to 10 million public alerts per night once the Vera C. Rubin Observatory’s Legacy Survey of Space and Time comes online, overwhelming the traditional physics-based inference pipelines. A continuous-time forecasting AI model is of interest because it can deliver millisecond-scale inference for thousands of objects per day, whereas legacy MCMC codes need hours per object. In this paper, we propose SELDON, a new continuous-time variational autoencoder for panels of sparse and irregularly time-sampled (gappy) astrophysical light curves that are nonstationary, heteroscedastic, and inherently dependent. SELDON combines a masked GRU-ODE encoder with a latent neural ODE propagator and an interpretable Gaussian-basis decoder. The encoder learns to summarize panels of imbalanced and correlated data even when only a handful of points are observed. The neural ODE then integrates this hidden state forward in continuous time, extrapolating to future unseen epochs. This extrapolated time series is further encoded by deep sets to a latent distribution that is decoded to a weighted sum of Gaussian basis functions, the parameters of which are physically meaningful. Such parameters (e.g., rise time, decay rate, peak flux) directly drive downstream prioritization of spectroscopic follow-up for astrophysical surveys. Beyond astronomy, the architecture of SELDON offers a generic recipe for interpretable and continuous-time sequence modeling in any time domain where data are multivariate, sparse, heteroscedastic, and irregularly spaced.
\end{abstract}
\vspace{6pt}
\noindent\textbf{Code:} \url{https://github.com/skai-institute/seldon}
\vspace{6pt}

\section{Introduction}

%\begin{itemize}
    %\item The discovery rate of optical transients will explode to $\sim$10 million public alerts per night once the Rubin Observatory’s LSST comes online, overwhelming the traditional physics-based inference pipelines. A foundation AI model is of interest because it can deliver millisecond-scale inference for hundreds of objects per day, whereas legacy MCMC codes need hours per object.
    %\item Optimization of spectroscopic followup is critical for next generation surveys such as Rubin.
    %\item This requires light curve extrapolation and prediction from few samples at early times.
    %\item Latent representations will generalize to broader tasks. In particular, it flags outlier-events for which no physical models exist, enabling real time discovery of ``unknown unknowns".
The arrival of the Rubin Observatory's Legacy Survey of Space and Time (LSST) will transform time-domain astronomy into a data-deluge era as it is expected to issue $\sim$10 million public alerts per night \citep{2019ApJ...873..111I}, overwhelming the capacity of traditional physics-driven analysis codes that rely on hours-long Markov-chain Monte Carlo (MCMC) runs per source. To support scientific discovery and maximize the return on limited spectroscopic resources, a new continuous-time forecasting AI model that can deliver millisecond-scale inference for hundreds of objects per day and extrapolate light curves from early, partial observations is urgently needed. 

Modeling astronomical time-domain light curves presents fundamental challenges. These nonstationary time series are sparse (often containing few measurements per object), heteroscedastic (exhibiting varying uncertainties across epochs), limited (covering only part of the event evolution), and irregularly spaced in time \citep{2022ApJS..258....1B}. Classic ARMA/ARIMA families are designed for evenly spaced and inherently stationary time series, and treating Rubin’s highly irregular cadence as missing data degrades both statistical power and interoperability. Also, their i.i.d. Gaussian-error assumption ignores the band-dependent photometric uncertainties, violating homoscedasticity assumptions and biasing parameter estimates \citep{feigelson2018arima}. Even when continuous-time generalizations (e.g., CARMA/CARFIMA) are used, inference scales cubically in the number of points, so a single supernova light curve can still take minutes to hours -- orders of magnitude slower than the millisecond budget required for $\sim$10 million alerts-per-night \citep{kelly2014carma}. There has been a growing body of work applying deep learning to transient light curves, but most existing models target either classification or coarse parameter regression rather than full multiband flux forecasting. For example, \textsc{SuperNNova} employs recurrent neural networks to assign supernova subtypes from partial photometry \citep{moller2019supernnova}, while \textsc{RAPID} leverages a recurrent neural network architecture to deliver near-real-time supernova type and rough peak-epoch estimates \citep{muthukrishna2019rapid}. \textsc{PELICAN} augments sparse LSST-like sequences with domain adaptation to improve classification under distribution shift \citep{pasquet2019pelican}, and \textsc{ORACLE} extends this line to a hierarchical, broker-scale classifier aimed at Rubin alert streams \citep{shah2025oracle}.

Autoencoders offer a powerful framework for learning compact task-agnostic representations for complex data, and the autoencoder families that operate on a fixed temporal grid have
long been the default solution for sequence modeling. Early examples such as STORN \citep{bayer2014storn}, VRNN \citep{chung2015vrnn},
SRNN \citep{fraccaro2016srnn}, and the Deep Kalman
Filter \citep{krishnan2017comp} combine a recurrent (or state–space) encoder with a generative decoder, delivering strong likelihoods on densely sampled videos and speech. More sophisticated variants embed a Gaussian-process prior in latent space \citep{fortuin2019gpvae}
or add score-based time extrapolation \citep{toth2020timegrad}, but all of these methods assume an equispaced input grid. When confronted with gappy, irregularly-spaced, and highly heteroscedastic Rubin-like light curves, they fail to make accurate  predictions on critical quantities such as peak time and flux.

Continuous-time latent models offer a principled route around the fixed-grid limitation. The seminal Neural ODE of \citet{chen2018neuralode} replaces the discrete RNN update with a differential flow $\dot z = f_\theta(z,t)$, allowing predictions at arbitrary time stamps.
From the encoder side, variants such as ODE–RNN \citep{rubanova2019latent} and GRU-ODE-Bayes \citep{li2020gruodebayes} integrate this flow only between
arrivals and therefore cope well with sparsity, but they decode with an unconstrained multilayer perceptron. In contrast, Latent ODE \citep{chen2018neuralode} and ODE2VAE \citep{yildiz2019ode2vae} generate directly in continuous time, yet still lack
band-specific, physically interpretable outputs. 
We combine the strength of previous work and provide SELDON, a novel architecture that marries a gap-aware encoder with a continuous-time latent propagator, condenses the resulting trajectory into a fixed-length summary, and channels that summary through an analytic decoder whose parameters are directly interpretable.
More specifically, our architecture has a masked GRU-ODE encoder that processes the sparse, irregularly-spaced, and heteroscedastic time series. At each observation, it performs a GRU update, then continuously propagates the hidden state between observations with a neural ODE, advancing the hidden state to the time of the next observation in the time series. The encoder's final latent vector serves as the initial condition for a downstream hidden ODE. Then, a continuous-time flow $\dot z = f_\theta(z,t)$ evolves an initial latent vector forward to a static regularly sampled grid, producing a dense trajectory $\{h(t_j)\}$. The trajectory is passed through a network $\phi$ and aggregated by a permutation-invariant pooling $\rho$, leading to a fixed-length representation $z = \rho(\{\phi(h(t_j))\})$. A \textsc{ResNet} takes $z$, together with a learnable embedding and returns for each band amplitudes, centers, and (inverse) widths of $K$ Gaussian basis functions. These Gaussian bases can deliver both pointwise predictions and inference about physically interpretable light curve attributes such as rise time, decay rate, peak time and flux that downstream schedulers use to prioritize scarce spectroscopic resources for followup in real time.
%\end{itemize}

\begin{figure}[t]
 \centering
\includegraphics[trim={3.8cm 0 1.5cm 0},clip,width=0.95\linewidth]{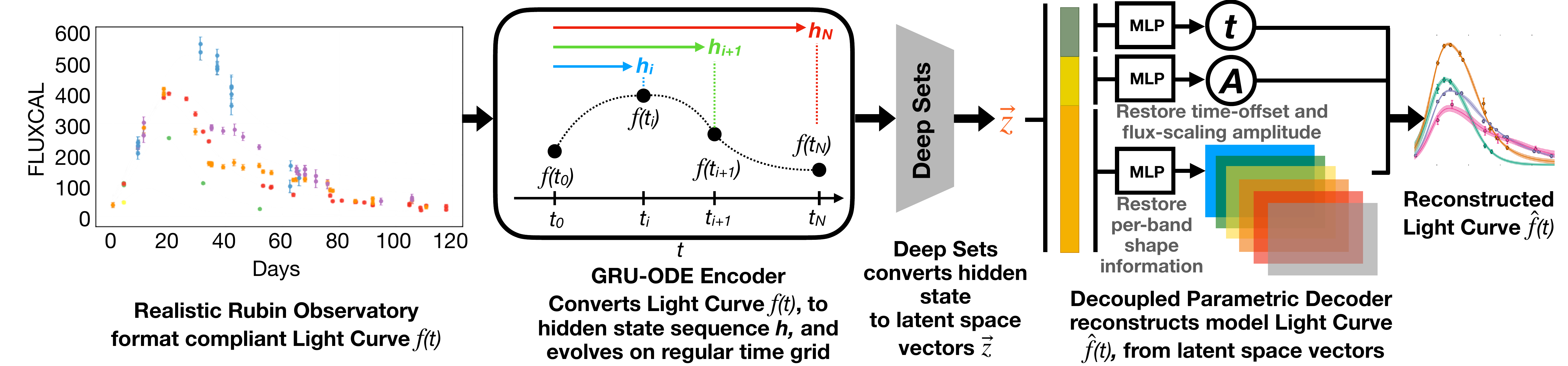}
 \caption{Architecture of our proposed SELDON, a customized VAE with band-aware GRU-ODE encoder and interpretable Gaussian-basis decoder.  A light curve described by a series of flux observations in various filter bands is encoded to an initial hidden state with the GRU-ODE.  The hidden state is evolved with the neural ODE forward in time to form a trajectory on a regularly-sampled grid.  This trajectory is then interpreted by a Deep Sets layer to an approximate posterior latent vector.  The latent vector is then decoded into a series of basis function parameters representing the history and future evolution of the light curve at all times in all filter bands. }
 \label{label:diagram}
\end{figure}

\section{Methods}

\subsection{Data}
Our work is driven by a new, publicly released data source, the \texttt{ELAsTiCC} astronomical multivariate time-series dataset \citep{2023AAS...24111701N}, hosted in {\texttt{Astro-ORACLE}} \citep{2025arXiv250101496S}. The library realistically simulated transients into a single HDF5 file with the canonical schema \texttt{\{MJD, band, FLUXCAL, FLUXCALERR\}}, representing the time of observation, the photometric filter with which the observation was taken, the calibrated flux in a given filter band, and the associated error with that flux measurement respectively. Focusing on the Type~Ia supernova (SN~Ia) class of transient events, every event is a multi-band ($\mathrm{u,g,r,i,z,y}$) photometric time series. For an overview of supernovae physics and evolution, see \citet{2017hsn..book.....A}.

We represent these data as panels of dependent time-domain light curves $LC_j$, such that each light curve contains a total of $N_j$ flux observations denoted by $f_{j, t_i, b_i}$ where $b_i$ is the photometric filter band at time point $t_i$, where $i$ indicates the $i^\text{th}$ observation in the light curve. 

\begin{figure}
    \centering
    \includegraphics[trim={0.5cm 14cm 0.0cm 14cm},clip,width=0.65\linewidth]{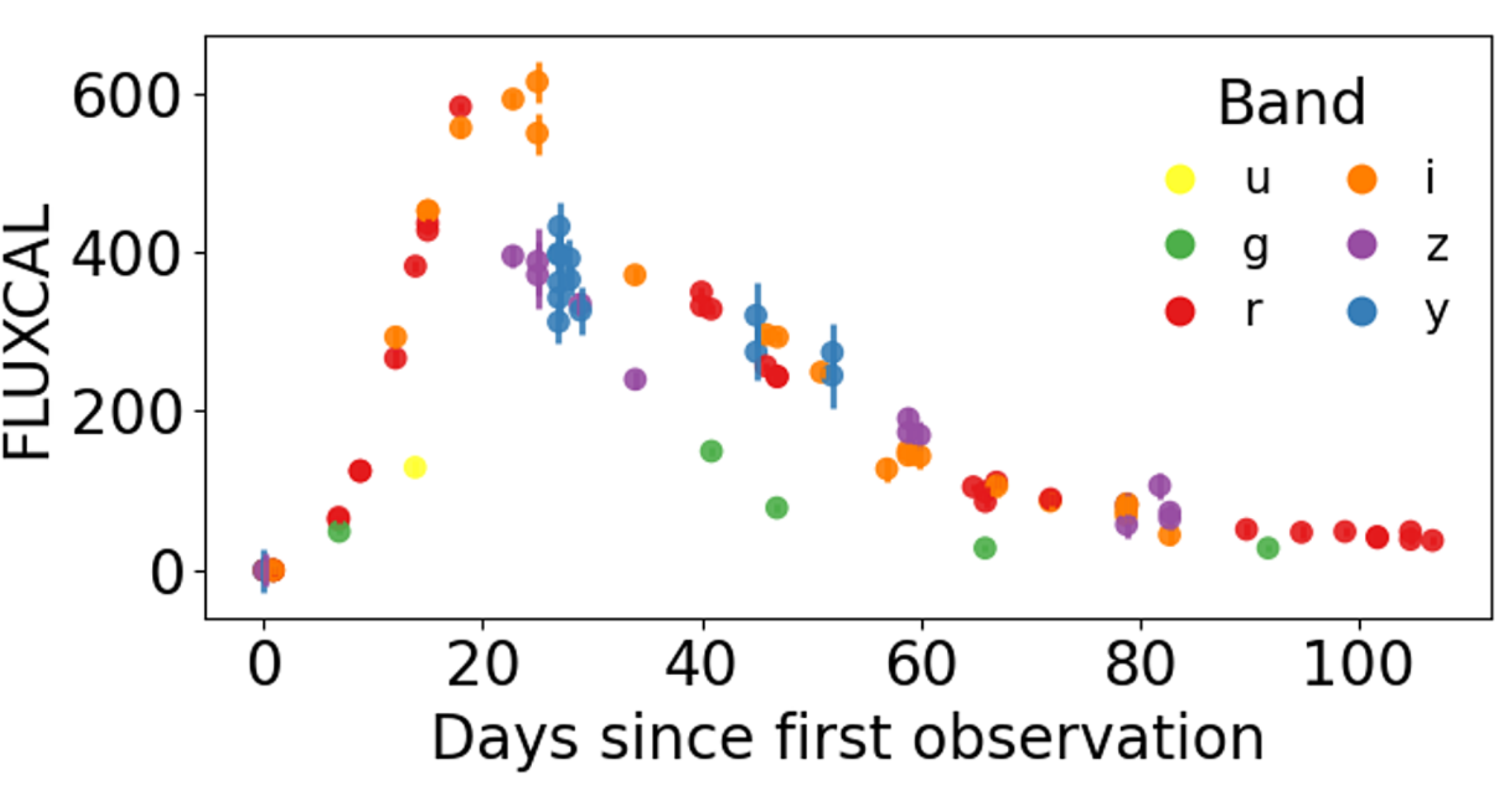}
    \caption{An illustration of a light curve observed over time across six bands indicated in distinct colors, where the total number of observations is in the $99^\text{th}$ percentile of all light curves. The error bars for each observation represent the observed flux errors.}
    \label{figure:lightcurve}
\end{figure}

Due to the fact that filters are  changed between observations, we can only observe the flux in a single filter band, at any given time for a single supernova. 
The data are irregularly spaced (gappy), sparse with few observations, and limited in that they often cover only part of the evolution of the light curve. Figure~\ref{figure:lightcurve} is an illustration of a light curve observed over time across six bands, where the total number of observations is in the $99^\text{th}$ percentile of all light curves. Typically, the total number of observations per light curves across all bands have a mean of 18 and a median of 14. Each point in a light curve has an observed flux error indicated as an error bar in Figure~\ref{figure:lightcurve}.

LSST is more sensitive in some filter bands (e.g., bands $\mathrm{g}$ and $\mathrm{r}$) than others (e.g., $\mathrm{u}$ and $\mathrm{y}$)~\citep{2008SPIE.7018E..2GO}, resulting in an imbalance of the data -- biased towards the more sensitive bands -- and varying signal-to-noise ratio between bands. 
In addition, data corresponding to each light curve are correlated within and between bands \citep{1997ARA&A..35..309F}.

These irregularly-sampled, inherently-dependent light curves are nonstationary, heteroscedastic, and nonlinear (Figure~\ref{figure:lightcurve}), thus posing substantial challenges in 
performing real-time prediction of light curves per band, in addition to recovering information across bands.

\subsubsection{Preprocessing}
For each light curve, we set the temporal origin of every light curve as follows. If an observed flux is below the acceptable signal-to-noise threshold for the survey, this observation is labeled as a non-detection point. Some non-detection points may have a negative flux due to systematic errors from background subtraction, and hence are noise dominated.  The flux of every non-detection point is then fixed at zero and its observed flux error remains unchanged.  
For each light curve, all non-detection points are omitted from our data except for up to 8 non-detection points immediately preceding the first detection point. This is needed to gain information about the point in time at which the light curve begins to rise.  The upper limit of  8 non-detection points is set at the median number of such points across all light curves.
We take the first of those non-detections as $t_0$. If there are no non-detections prior to the detection point, we set $t_0$ to be the time of the first detection point which coincides with the discovery time at which we observe the first detection point. 
Then, we rescale time by $t_{\mathrm{norm}}=2\sigma$, where $\sigma$ is the standard deviation of full-curve durations over the training set. The resulting $\tilde t_i = (t_i - t_0)/t_{\mathrm{norm}}$ keeps gradient magnitudes well scaled.  From the training data, we set $t_{\mathrm{norm}}$ to be \num{71.9}\,days.

Raw flux measurements span several orders of magnitude and carry heteroscedastic errors. 
To bring all measurements onto a numerically stable scale for training purposes, 
we apply a signed logarithmic compression followed by symmetric scaling. 
Specifically, we use the log-modulus transformation
\begin{equation}
g(f)=\operatorname{sgn}(f)\,\log_{10}\bigl(|f|+1\bigr), 
\qquad 
\sigma_g=\frac{\sigma_f}{(|f|+1)\,\ln 10}.
\end{equation}
This transformation makes the flux distribution approximately symmetric 
and keeps magnitudes $\mathcal{O}(1)$ for stable optimization. 
We then scale $(g,\sigma_g)$ by the dataset-wide maximum absolute value 
$g_{\max}=\max_{\text{train}}|g|$ and recenter
\begin{equation}
\tilde g = \frac{g}{g_{\max}} - 0.5.
\end{equation}
In practice this maps $\tilde g$ close to $[-0.5,0.5]$ while preserving relative magnitudes. 
The inverse transform uses $g = (\tilde g + 0.5) g_{\max}$ followed by $f = \operatorname{sgn}(g)\,(10^{|g|}-1)$.

\subsubsection{Augmentation}
At each training step, the model is trained on a partial light curve, which is a freshly generated, incomplete version of full light curve. More precisely, we keep only the first $K$ measurements, where $K$ is drawn uniformly between a minimum of 10 points and the full length of the light curve. Because $K$ is almost always smaller than the phase of maximum flux, these cut-offs mainly contain the rising part of the light curve, which is exactly the scenario faced by survey schedulers who must decide follow-up strategy before the peak. By generating these partial curves on the fly, the model encounters a new mix of truncated curves in every mini-batch and learns to make reliable predictions from whatever segment of the light curve it happens to receive.

\subsection{Architecture}
We develop a customized variational autoencoder (VAE) in which a band-aware GRU-ODE encoder with deep sets maps each partial light curve to a latent Gaussian  $\mathcal N(\boldsymbol\mu,\operatorname{diag}\boldsymbol\sigma^2)$ of dimension $D_z$ (we use $D_z=64$), and a parametric basis decoder reconstructs flux values on any provided continuous time grid from a linear combination of Gaussian basis functions. Figure~\ref{label:diagram}  provides an overview of our developed architecture pipeline. Below we describe the various components of our proposed VAE in detail.  Hyperparameter values for the models described below are shown in Table~\ref{tab:hp}.

\subsubsection{Embedding}
Every measurement is represented as a fixed-width six-channel vector 
$[\tilde t_i,\,\tilde g_i,\,\mathbf e_i^\top]^\top$, 
where $\tilde t_i$ is the normalized time, 
$\tilde g_i$ is the log-scaled flux after rescaling, 
and $\mathbf e_i\in\mathbb{R}^4$ is a learnable band embedding. 
The band information is essential for capturing the color evolution of the supernovae. 
Each photometric band $\mathrm{u,g,r,i,z,y}$ is assigned an integer index that 
retrieves $\mathbf e_i$ from an embedding matrix 
$\mathbf E\in\mathbb{R}^{6\times4}$ initialized randomly and learned jointly with the network. 
To mitigate band-frequency imbalance, gradients of each embedding vector are scaled 
by the inverse occurrence frequency of its band by dividing the gradients in each embedding by the number of corresponding samples in each mini-batch. 
Thus every observation is encoded as 
\[
\mathbf z_i = [\tilde t_i,\,\tilde g_i,\,\mathbf e_i^\top]^\top \in \mathbb{R}^6,
\]
providing a compact representation that combines time, flux, and band information 
for input to the encoder.

\subsubsection{Encoder}
The encoder takes a sparse irregular light curve and returns a
Gaussian posterior over an initial latent state $z_0$. We implement
three competing variants: a traditional GRU encoder \citep{cho-etal-2014-properties}, a permutation-invariant Deep Sets network model \citep{2017arXiv170306114Z}, and a GRU-ODE network model \citep{2019arXiv190703907R} with Deep Sets.

\subsubsection{Gated Recurrent Unit}
The GRU is a lightweight gated-RNN architecture proposed to mitigate the vanishing or exploding gradient problems that plague vanilla RNNs. We consider a single-layer GRU with \texttt{hidden\_dim} hidden units that processes the six-dimensional input $[\tilde t_i, \tilde g_i, \mathbf e_{i}^{\top}]^{\top}$ and outputs a \texttt{hidden\_dim}-dimensional hidden state that memorizes the entire light curve for the downstream decoder.

\subsubsection{Deep Sets}
Deep Sets provides a simple way to encode a variable-length order-agnostic collection of items.
To implement it, we use an element-wise network $\phi$ that maps each observation to
a \texttt{hidden\_dim}-dimensional space, followed by a sum pooling,
$\sum_t\phi(\mathbf x_t)$, and a second MLP $\rho$ outputting $\boldsymbol\mu$ and $\log\boldsymbol\sigma^{2}$.
This encoder is permutation invariant and preserves the association between flux observations and time as they are initially passed as pairs to this encoder.

\subsubsection{GRU-ODE Encoder with Deep Sets}
A single-layer, unidirectional GRU with \texttt{hidden\_dim} hidden units 
takes five-dimensional inputs 
$[\tilde g_i,\,\mathbf e_i^\top]^\top$ 
in reverse chronological order, omitting time since temporal evolution 
is handled explicitly by the continuous ODE between observations. 
Between observation times, the hidden state $h(\tilde t)$ is 
propagated by an autonomous latent ODE
\begin{equation}
\frac{\mathrm{d}h_{\tilde t}}{\mathrm{d}\tilde t}=f_\theta(h_{\tilde t}),
\end{equation}
and updated by the GRU at each measurement, ensuring smooth trajectories 
for irregularly spaced light curves. 
The final hidden state serves as the initial condition for a forward ODE, 
which evolves this state on a regularly sampled time grid of \num{50} points 
for $\tilde t\!\in\![0,1]$, producing a 
\texttt{hidden\_dim}$\times$\num{50} hidden-state trajectory. This corresponds to a $\approx$72 day evolution which adequately encompasses the evolution of SN~Ia optical light curves.
This trajectory is concatenated with the corresponding time grid and 
encoded by a \texttt{Deep Sets} module with sum aggregation to yield the 
approximate posterior
\[
q_\phi(\mathbf z\mid\mathbf x)
=\mathcal N(\boldsymbol\mu,\operatorname{diag}\boldsymbol\sigma^{2}).
\]

\subsubsection{Latent Neural-ODE Solver}
We integrate the latent flow $\dot z = f_{\theta}(z)$ with the adaptive \textsc{Tsit5} solver from \textsc{TorchODE} \citep{lienen2023torchode} with a maximum step size $dt$ of 0.01 in normalized time space.
The entire solver is wrapped in an \textsc{ODESolver} module that is
\texttt{torch.compile}-optimized, giving a $\sim\!2\times$ speed-up over the default eager-mode PyTorch execution. 
At run time, we can switch between the
\emph{AutoDiffAdjoint} path, which stores the forward trajectory and is fastest when memory allows, and the memory-lean
\emph{BacksolveAdjoint} path, which recomputes states during the
backward sweep.
Both adjoint choices deliver gradients accurate up to solver tolerances and discretization error, and the solver naturally parallelizes over a batch of latent trajectories, allowing us to propagate hundreds of light curves per second on a single GPU.

\subsubsection{Decoder}

During training we use the re-parameterization trick,
drawing
$\mathbf z=\boldsymbol\mu+\boldsymbol\sigma\odot\boldsymbol\epsilon$
with $\boldsymbol\epsilon\sim\mathcal N(\mathbf 0,\mathbf I)$ \citep{kingma2014autoencoding}.  
At test time, we set $\mathbf z=\boldsymbol\mu$ to obtain a single 
deterministic reconstruction. We use a parametric Gaussian basis decoder, for which the flux in band $b$ is modeled as a sum of
  $K$ Gaussian basis functions 
\begin{equation}\label{eq:gauss}
  \hat f_b(t)=\sum_{k=1}^{K} w_{bk}
  \exp\left[-\left((t-\mu_{bk})\sigma_{bk}\right)^{2}\right].
\end{equation}
In \eqref{eq:gauss}, for each band $b$ and basis component $k$, the decoder predicts three parameters: the amplitude $w_{b,k}$, the center time $\mu_{b,k}$, and the rate $\sigma_{b,k}$ that controls its spread. Together, these parameters let the sum of $K$ Gaussians flexibly trace the rise and fall of the light curve in band $b$.

The Gaussian basis decoder takes
$[\mathbf z]$, passing it through a four-layer
\textsc{ResNet} with hidden size \texttt{hidden\_dim}, and outputs the parameters of a
band–specific analytic light-curve model that can be evaluated on any
query grid $\{t_j\}$. We multiply the latent
vector of basis function parameters element-wise with the band embedding and sum them to a set of $K=8$ basis parameters for each band, reducing the input dimensionality without halting gradient flow. The MLP weights are shared across bands, while
band-to-band variation enters through another learned embedding
$\mathbf e_{b,decoder}$ with \num{16} output dimensions.

Amplitudes and centroids are
mapped through a learnable exponential or normal inverse transform mapping, respectively.  These enforce the parameter distributions to be well behaved during early training epochs.  By initializing the learnable parameters of these mappings, we tune the network to make initial predictions close to reasonable values for these parameters. 
We scale the amplitudes by dividing each amplitude by the mean of the set of $K$ amplitudes.  We center the centroids by subtracting the mean of the set of $K$ centroids from each centroid.

A global amplitude and centroid are then decoded from a decoupled segment of the latent vector, each from their own independent \textsc{ResNet}, and projected onto their own respective inverse transform sampler.  The individual basis amplitudes are then multiplied by the global amplitude, and the global centroid is added to the individual centroids.  Of the 64 components of the latent vector, the individual basis parameters are decoded from the first 48 components.  The global centroid is decoded from the next 8 components of the latent vector, and the global amplitude is decoded from the final 8 components of the latent vector.  The independence of the global parameters allows for a decoupling between the overall amplitude of the light curve and its time offset, providing scale and time invariance to the individual components.

\subsubsection{Learnable Inverse Transform Mapping}
We use a learnable distribution mapping for certain basis parameters in order to improve training performance by enforcing a restricted initial distribution for early epochs which is then relaxed during training.
Particularly, an inverse transform is used to learn the parameters $\mu$ and $\sigma$ of a normal distribution applied to the individual and global centroids with initial respective values of 0.0 and 2.0 for $\mu$ and an initial value of 0.1 for $\sigma$.
For the case of individual and global amplitudes, an exponential distribution is applied with the initial values for $\lambda$ set to 0.25 and 8.0, respectively.

\paragraph{Loss}
For each light curve $LC_j$ observed at time $t_{i}$, let
$\hat f_{j,t_i,b_i}$ be the reconstructed flux, $f_{j,t_i,b_i}$ the observed
flux, and $\sigma_{j,t_i,b_i}$ the reported flux error.
A binary mask $m_{i}\in\{0,1\}$ identifies valid (unpadded) samples at the $i^\text{th}$ observation in the light curve.
With numerical safeguard $\varepsilon=10^{-6}$ and Huber scale
$\delta=1$, we apply Huber loss $\ell_{j,t_i,b_i}$ to the standardized residual 
$
r_{j,t_i,b_i}= \frac{f_{j,t_i,b_i}-\hat f_{j,t_i,b_i}}{\max(\sigma_{j,t_i,b_i},\varepsilon)}.
$

\begin{comment}

The per timestamp robust penalty is
\begin{equation}
\ell_{t_i,b_i}=
\begin{cases}
\dfrac{1}{2}|r_{t_i,b_i}|^{2}, & |r_{t_i,b_i}|<\delta,\\[6pt]
\delta\left(|r_{t_i,b_i}|-\dfrac{\delta}{2}\right), & |r_{t_i,b_i}|\ge \delta.
\end{cases}
\end{equation}
\end{comment}

Masking out padded entries and averaging over the valid points of each
sequence yields the per-curve loss
\begin{equation}
\mathcal{L}_{\mathrm{rec},j}=
\frac{\displaystyle\sum_{i} m_{j,i}\,\ell_{j,t_i,b_i}}
     {\displaystyle\sum_{i} m_{j,i}+\varepsilon}.
\end{equation}
%The scalar training objective is the mean over a batch of size $B$,
%$\mathcal{L}=\tfrac{1}{B}\sum_{j=1}^{B}\mathcal{L}_{j}$, providing a
%robust criterion that is agnostic to sequence padding.  
To regularize the latent space, we add the Kullback–Leibler divergence
between the approximate posterior
$q_\phi(\mathbf z\mid \mathbf x)\sim\mathcal N(\boldsymbol\mu,\operatorname{diag}\boldsymbol\sigma^2)$
and the unit Gaussian prior $p(\mathbf z)\sim\mathcal N(\mathbf 0,\mathbf I)$ given by
\begin{equation}
D_{\mathrm{KL}}\big(q_\phi(\mathbf z\mid\mathbf x)\,\|\,p(\mathbf z)\big)
=\tfrac{1}{2}\sum_{d=1}^{D_z}\big(\mu_d^2+\sigma_d^2-\log\sigma_d^2-1\big),
\end{equation}
where $D_z$ is the latent dimensionality (set to 64 in our experiments).
The total per-curve objective is therefore
\begin{equation}
\mathcal{L}_j = \mathcal{L}_{\mathrm{rec},j} + \beta\,D_{\mathrm{KL}}\!\big(q_\phi(\mathbf z\mid\mathbf x)\,\|\,p(\mathbf z)\big),
\end{equation}
with $\beta=10^{-4}$ used as a fixed constant throughout training.
The scalar training objective is the mean over a batch of size $B$, given by
\begin{equation}
\mathcal{L}
=\frac{1}{B}\sum_{j=1}^{B}\mathcal{L}_j.
\end{equation}
This formulation combines a robust, uncertainty-aware reconstruction term
with a variational regularizer, yielding stable training and a well-behaved latent prior.

\begin{figure}[H]
    \centering
    \includegraphics[trim={0.0cm 12cm 0.0cm 12cm},clip,width=0.32\linewidth]{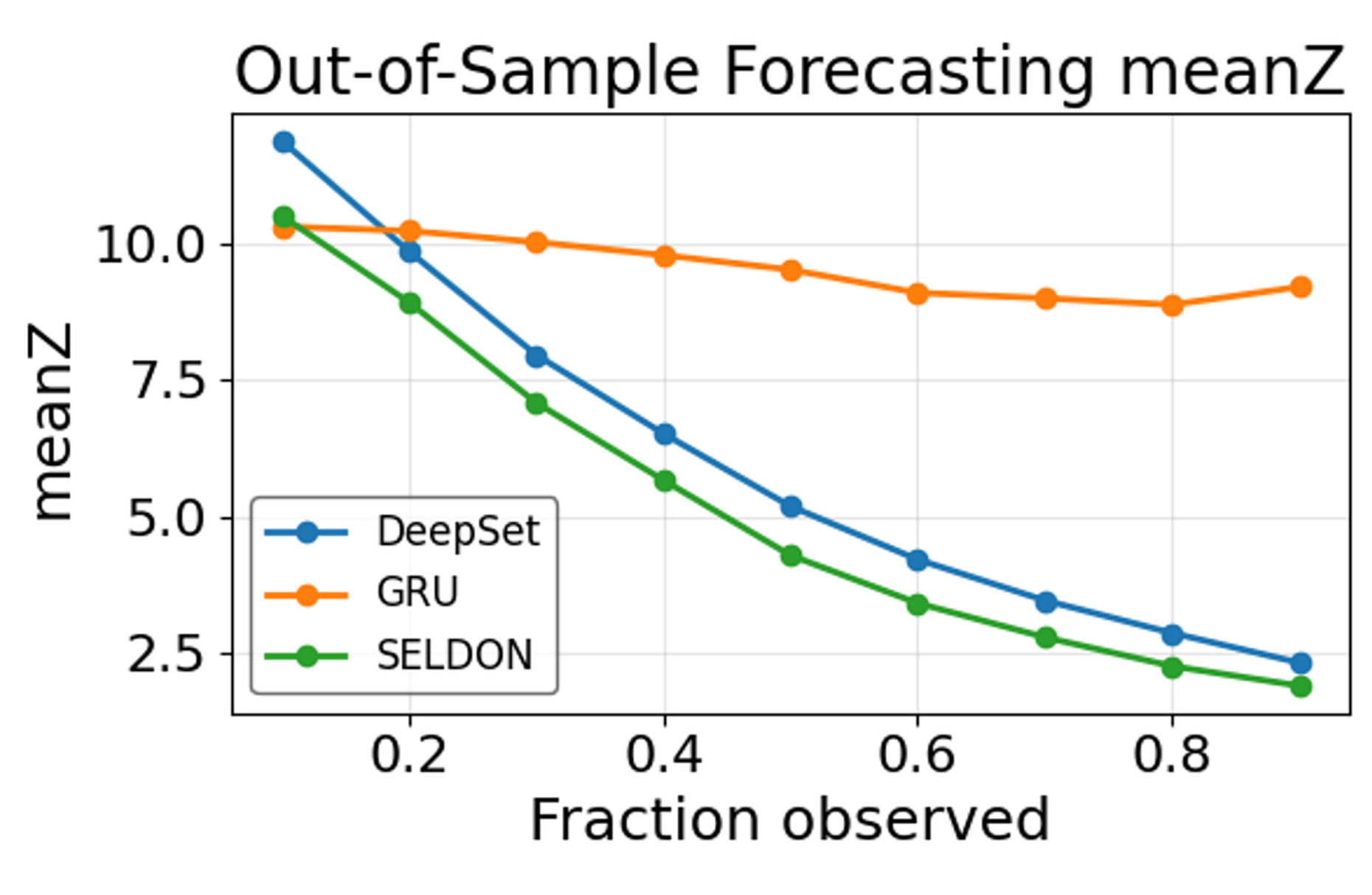}\hfill
    \includegraphics[trim={0.0cm 12cm 0.0cm 12cm},clip,width=0.32\linewidth]{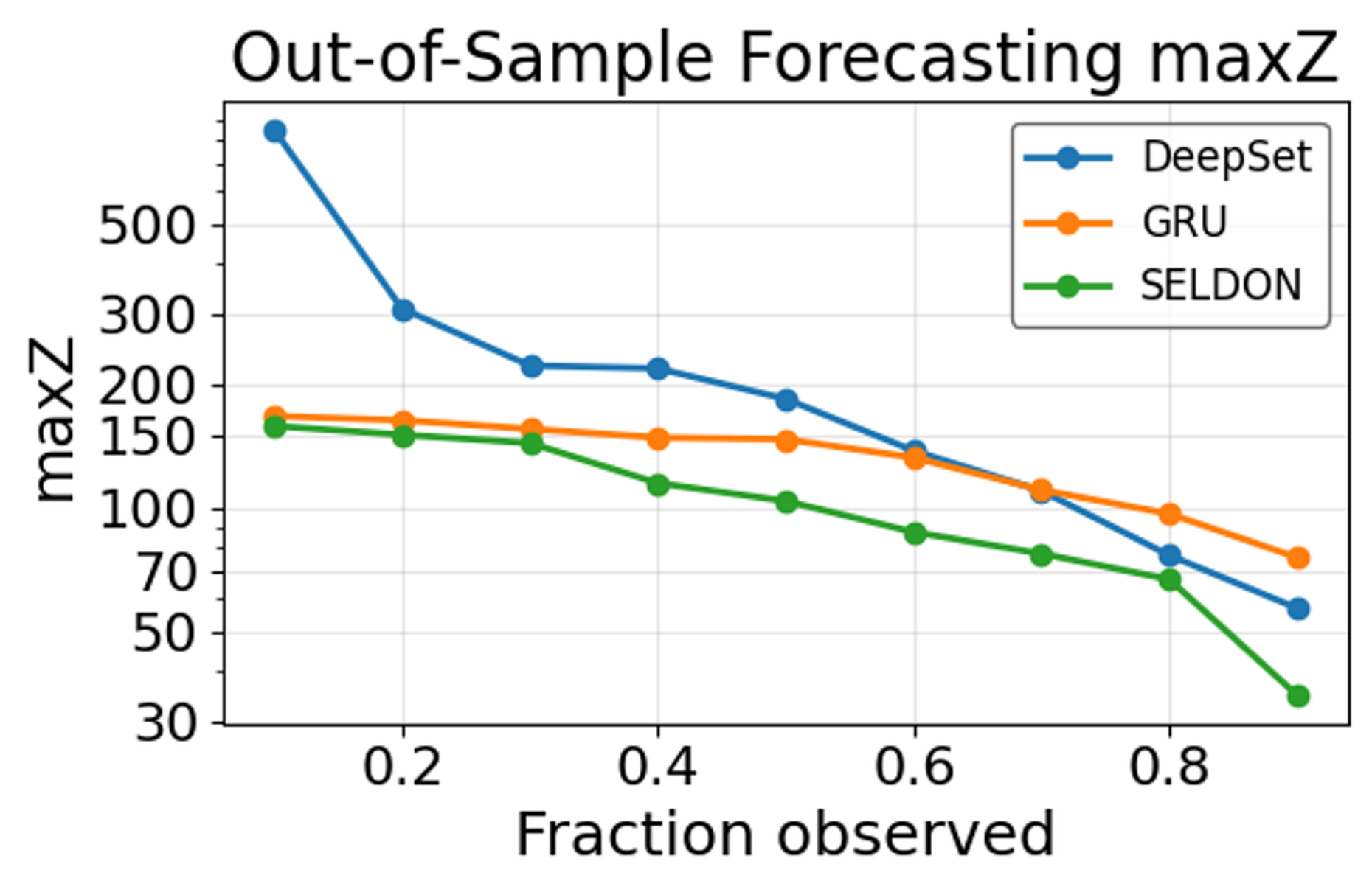}\hfill
    \includegraphics[trim={0.0cm 12cm 0.0cm 12cm},clip,width=0.32\linewidth]{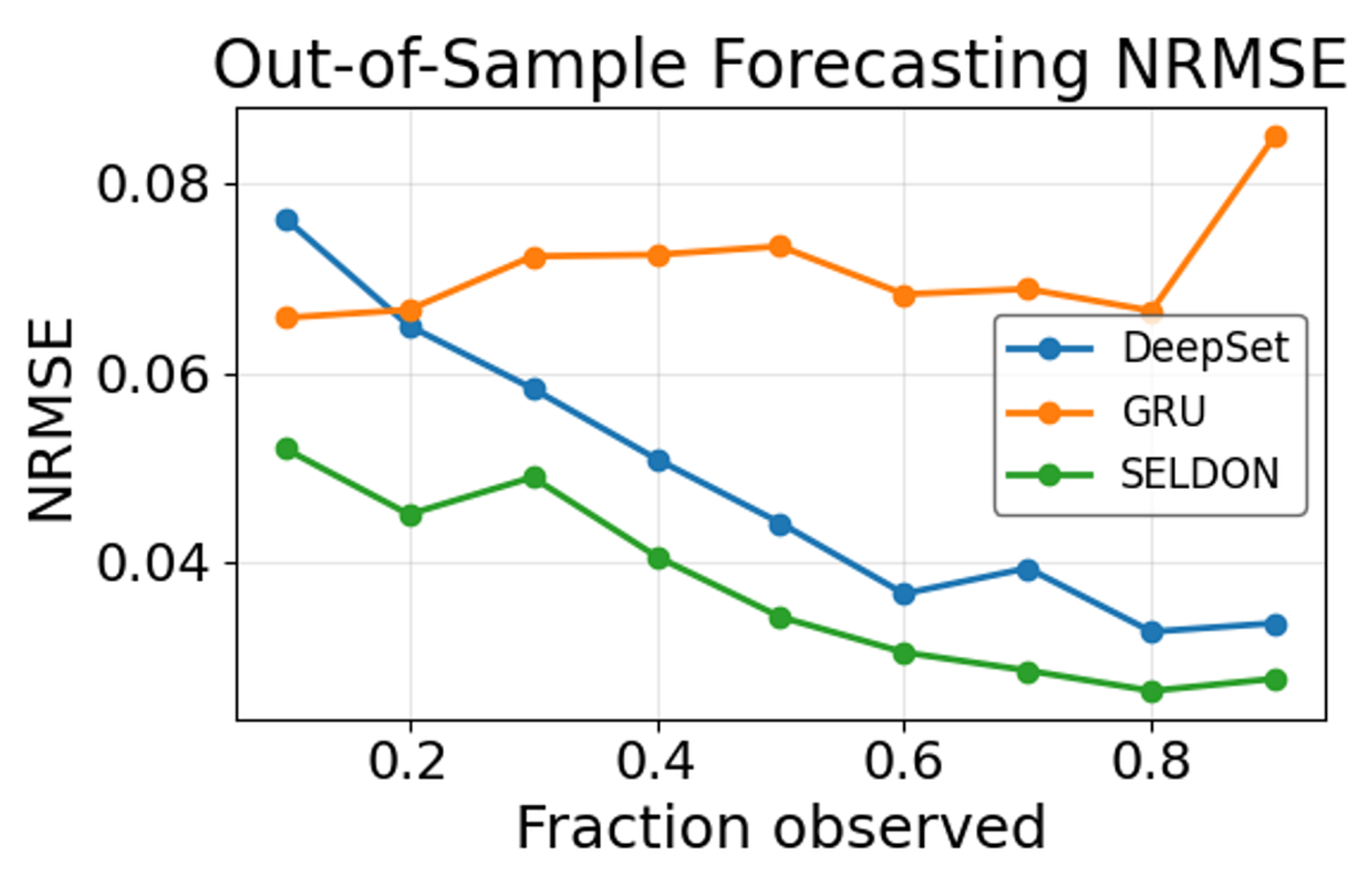}
    \caption{Out-of-sample forecasting performance as a function of the fraction of the
    light curve that has been observed. \textbf{Left}: mean absolute $Z$-score ($\mathrm{mean}\,|Z|$).
    \textbf{Center}: worst-case absolute $Z$-score ($\mathrm{max}\,|Z|$, log-scale).
    \textbf{Right}: normalised RMSE. Lower is better in all panels. SELDON (i.e., GRU-ODE) (green) consistently produces the lowest tail and aggregate errors.
    A plain masked-GRU (orange) has the best median at $10\%$ observed but is outperformed by SELDON afterward.
    Deep Sets (blue) shows competitive medians, yet the heaviest tails.}
    \label{fig:metrics_plots}
\end{figure}

\begin{table}
    \centering
    \begin{tabular}{c|c|c|c}
       Model&Masked-GRU& Deep Sets & SELDON \\
       \hline
       \texttt{hidden\_dim}  & 256 & 256 & 128\\
       Learning Rate & 0.0001 & 0.002 & 0.002\\
       Batch Size & 2048 & 2048 & 2048 \\

    \end{tabular}
    \caption{Hyperparameters for the three encoder models compared within this work. Hyperparameters were optimized through grid search.}
    \label{tab:hp}
\end{table}

\subsubsection{Training}

SELDON was trained for 180 epochs on a Nvidia H100 GPU using the \texttt{Adam} optimizer with $\beta_{1}$ = 0.9, $\beta_{2}$ = 0.999, and $\epsilon$ = $10^{-8}$.  A single training step took 7.5 seconds with 1.1 seconds for inference.  Training used batch accumulation over 4 mini-batches of size 512 and required 45GB of GPU memory.  The model was trained until the validation loss stopped improving.
%\end{comment}

\section{Results}

\subsubsection{Performance metrics}
We adopt the following three metrics to evaluate the model forecasting performance of a panel of inherently-dependent time-domain light curves $LC_j$, such that each light curve contains a total of $N_j$ flux predictions given by  $\hat{f}_{j, t_i, b_i}$, with observed flux $f_{j, t_i, b_i}$ and observed flux error $\sigma_{j, t_i, b_i}.$ For simplicity, we denote the $i^\text{th}$ flux observation of  $LC_j$ by $f_{j,i}$ with observed flux error $\sigma_{j,i}$ and the associated predicted flux as $\hat{f}_{j,i}$. While we are using simpler notations for this section, it is important to note that the $i^\text{th}$ flux observation of a light curve $LC_j$ implicitly refers to a filter band $b_i$ and time point $t_i$.

\begin{comment}
We compute the absolute fractional error at the point $i$
\begin{equation}
\mathrm{FE}_{j, i}=\frac{\left|\hat{f}_{j, i}-f_{j, i}^{\text {true}}\right|}{\max \left(f_{j, i}^{\text {true}}, F_{\text {min}}\right)},
\end{equation}
where $F_{\text{min}}$ is the flux floor (e.g., 1) that prevents division by zero and is the same constant everywhere in the code. 
\end{comment}
Define the standardized $Z$-score for the $i^\text{th}$ flux observation of light curve $LC_j$  as follows
\begin{equation}
\left|Z_{j, i}\right|:=\frac{\left|f_{j, i}-\hat{f}_{j, i}\right|}{\sigma_{j, i}}. 
\end{equation}

Then, the mean absolute $Z$-score and the maximum absolute $Z$-score are defined as
\begin{equation}
\overline{|Z|}_j:=\frac{1}{N_j} \sum_{i=1}^{N_j}\left|Z_{j, i}\right|, \quad \operatorname{Max}|Z|_j:=\max _{1 \leq i \leq N_j}\left|Z_{j, i}\right|.
\end{equation}
\begin{comment}
We then have the per-curve aggregated metrics 
\begin{equation}
\mathrm{MeanFE}_j=\frac{1}{N_j} \sum_{i=1}^{N_j} \mathrm{FE}_{j, i},
\end{equation}
\begin{equation}
\operatorname{Max}_{\mathrm{FE}_j}=\max _{1 \leq i \leq N_j} \mathrm{FE}_{j, i},
\end{equation}
\end{comment}
Assuming the test set has $M$ curves, then the overall metrics are defined as 
\begin{comment}
%\begin{equation}
%\mathrm{MeanFE}^{\text {overall }}:=\frac{1}{M} %\sum_{j=1}^M \mathrm{MeanFE}_j,
%\end{equation}
%\begin{equation}
%\mathrm{MaxFE}^{\text {overall }}:=\frac{1}{M} %\sum_{j=1}^M \mathrm{MaxFE}_j,
%\end{equation}
\end{comment}
\begin{equation}
\text{Mean}|Z|:=\frac{1}{M} \sum_{j=1}^M \overline{|Z|}_j, \quad
\operatorname{Max}|Z| := \max_{1 \le j \le M} \operatorname{Max}|Z|_j,
\end{equation}

The normalized root mean-square error $\mathrm{NRMSE}$ is defined as
\begin{equation}
\mathrm{NRMSE} := \frac{1}{M} \sum_{j=1}^{M}
\frac{
\sqrt{\tfrac{1}{N_j}\sum_{i=1}^{N_j}\!\bigl(f_{j,i}-\hat{f}_{j,i}\bigr)^2}}
{\max_i f_{j,i}},
\end{equation}
which is the average of the root-mean-square reconstruction errors normalized by the curves'
maximum observed fluxes, yielding a dimensionless quantity.

\subsection{Forecasting Performance}

\begin{table}[H]
  \centering
  \small

  \setlength{\tabcolsep}{4.2pt}  
  \begin{tabular}{@{}l
                  rrr  % Mean|Z|
                  rrr  % Max|Z|
                  rrr@{}} % NRMSE
    \toprule
    & \multicolumn{3}{c}{Mean$|Z|$ $\ \downarrow$}
    & \multicolumn{3}{c}{Max$|Z|$ $\ \downarrow$}
    & \multicolumn{3}{c}{NRMSE $\ \downarrow$} \\
    \cmidrule(lr){2-4}\cmidrule(lr){5-7}\cmidrule(l){8-10}
    Fraction
      & Deep Sets & Masked-GRU & SELDON
      & Deep Sets & Masked-GRU & SELDON
      & Deep Sets & Masked-GRU & SELDON \\
    \midrule
    0.1 & 11.885 & \textbf{10.315} & 10.513
        & 848.772 & 168.496 & \textbf{159.292}
        & 0.076 & 0.066 & \textbf{0.052} \\
    0.2 &  9.862 & 10.237 & \textbf{8.929}
        & 309.186 & 164.536 & \textbf{151.551}
        & 0.065 & 0.067 & \textbf{0.045} \\
    0.3 &  7.966 & 10.031 & \textbf{7.089}
        & 224.164 & 156.688 & \textbf{144.667}
        & 0.058 & 0.072 & \textbf{0.049} \\
    0.4 &  6.525 &  9.792 & \textbf{5.673}
        & 220.531 & 149.150 & \textbf{115.359}
        & 0.051 & 0.072 & \textbf{0.041} \\
    0.5 &  5.193 &  9.526 & \textbf{4.295}
        & 185.677 & 147.793 & \textbf{104.237}
        & 0.044 & 0.073 & \textbf{0.034} \\
    0.6 &  4.218 &  9.103 & \textbf{3.418}
        & 138.518 & 133.256 & \textbf{87.474}
        & 0.037 & 0.068 & \textbf{0.031} \\
    0.7 &  3.469 &  9.002 & \textbf{2.794}
        & 110.469 & 110.968 & \textbf{77.439}
        & 0.039 & 0.069 & \textbf{0.029} \\
    0.8 &  2.872 &  8.884 & \textbf{2.266}
        &  76.708 &  96.979 & \textbf{66.972}
        & 0.033 & 0.067 & \textbf{0.027} \\
    0.9 &  2.329 &  9.209 & \textbf{1.906}
        &  56.891 &  75.786 & \textbf{34.789}
        & 0.034 & 0.085 & \textbf{0.028} \\
    \bottomrule
  \end{tabular}
    \caption{Out–of–sample performance per fraction of observed values. Down arrows next to the metrics  indicate lower values are better.   \textbf{Bold} values indicate best overall performance across all models.}
    \label{tab:all_metrics}
\end{table}

% Explain here the motivation for the metrics we use for forecasting, using percentiles of light curve seen

% Explain the plots, what they show, and why these are the best representations
We evaluate the performance of our proposed SELDON against the two other models with the masked-GRU encoder and the Deep Sets encoder, on out-of-sample predictions cutoff at a given percentage of sequential data points that are assumed to be observed. Table~\ref{tab:all_metrics} reports two sets of aggregate metrics divided into two categories, absolute and relative metrics.  
The results show that SELDON with Deep Sets model achieves the best performance of the three we have tested across all columns of Table~\ref{tab:all_metrics} (except at 10\% observed fraction where the Mean $|Z|$ of GRU is slightly better) and with increasing performance over other models at earlier times in the sequence. This is also illustrated in Figure~\ref{fig:metrics_plots}, which displays the out-of-sample forecasting performance as a function of the percentage observed which is consistently favoring our SELDON model against the masked-GRU model and the Deep Sets model.

\begin{figure}[H]
 \centering
\includegraphics[trim={0cm 18.3cm 0cm 18cm},clip,width=0.95\linewidth]{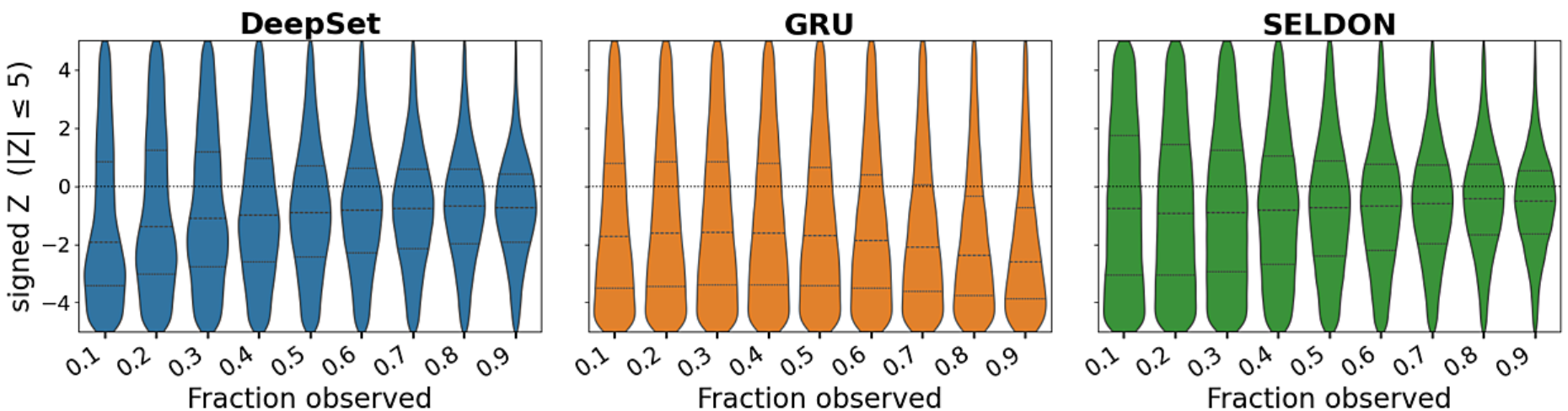}
\caption{Violin plots of the signed standardized residuals for out-of-sample forecasts per fraction observed, in each of the three models. These residuals are clipped between $\pm 5$ for visual clarity.}
\label{label:violin}
\end{figure}

Figure~\ref{label:violin} reports the violin plots of the signed standardized residuals for out-of-sample forecasts per percentage observed, in each of the three models. The violin plots demonstrate that the masked-GRU encoder has consistently poor performance regardless of the percentage of the light curve revealed. While the Deep Sets model has poor performance for low percentages, it does improve as more data is revealed though not as quickly as SELDON.  For our applications, it is most crucial to have accurate real-time predictions at the early stages of the time series before the peaks are observed, a task that is best achieved by our SELDON and consistently across all the metrics used.

When only $10\%$ of the light curve is revealed, the task is essentially extrapolation from a handful of early points. In that regime the masked-GRU model attains the smallest mean $|\mathrm{Z}|$ (i.e., $10.3 \sigma$), edging SELDON by $<2\%$. The difference disappears once $\geq 20\%$ of the curve is available: SELDON takes the lead and keeps it through 90\% coverage, cutting the average error by 10-35\% relative to GRU and by 15-45\% relative to Deep Sets. The gain grows with the fraction because the latent-ODE decoder can exploit every additional observation to tighten its continuous trajectory, whereas the discrete-time masked-GRU is limited to fixed-step updates.

The Max$|Z|$ highlights a contrast. Deep Sets occasionally produces catastrophic residuals, hundreds of $\sigma$ at $20\%$ coverage and nearly $900\sigma$ at $10\%$. A vanilla masked-GRU removes the worst spikes but still exhibits excursions above $160\sigma$. SELDON caps the worst error below $160\sigma$ in the sparsest slice and below $90\sigma$ thereafter, delivering the tightest upper tail throughout.

Every standardized residual in the NRMSE is weighted quadratically, and therefore it penalizes both bias and heavy tails. SELDON achieves the lowest NRMSE at every percentage of observed values, $20-35\%$ lower than masked-GRU and $30-50 \%$ lower than Deep Sets, showing that it not only reduces the mean error but also suppresses large deviations effectively.

%Overall, Deep Setss, which pools observations in a permutation-invariant way, does a good job 
%\begin{itemize} 
%    \item Light Curve peak predictions
%    \item Partial light-curve reconstruction
%\end{itemize}

%\subsection{Timing Benchmarks}

%\begin{tabular}{c|c|c}
%    Encoder &  Time per Sample & Peak Prediction Accuracy\\
%    \hline
%    GRU-ODE & ... &  ...\\
%    GRU & ... & ... \\
%    Deep Setss & ... & ...\\
%\end{tabular}

% Hyperparameters, activation functions, learning rates, dropouts, number of hidden layers

\section{Discussion}

We developed a forecasting VAE model for SNe~Ia optical light curve evolution based on a combination of GRU-ODE encoder and Deep Sets with a Gaussian basis decoder, accounting for band-frequency imbalance. In comparison to a pure masked-GRU encoder and a pure Deep Sets encoder, we find that our model demonstrates superior performance consistently across a variety of forecasting metrics.  The Gaussian basis decoder provides flexibility in forecasting by allowing evaluation to be performed at any point in time from a functional representation produced from encoded observations.  By decoupling global parameters of the basis function in the latent dimension, we provide local scale- and time-invariant intrinsic basis parameters describing the evolution of the supernova.  

SELDON is capable of handling sparse, heteroscedastic, band-frequency-imbalanced multivariate time series with irregular sampling. These panels of gappy time series are inherently dependent and exhibit a nonlinear and nonstationary behavior. We have shown that forecasting such data can be achieved using our GRU-ODE plus Deep Sets encoder model. In particular, SELDON has a unique crucial feature in predicting light curves in the early times before the peak is observed with exceptionally limited amount of data observed. Developing such a framework is useful in other fields of astronomy as well as those outside astronomy.

In the age of the Vera C. Rubin Observatory, SELDON is easily capable of keeping up with the expected LSST alert rate, helping provide predictions for optimal spectroscopic followup scheduling.

\section{Acknowledgments}
We gratefully acknowledge the support of the NSF-Simons AI-Institute for the Sky (SkAI) via grants NSF AST-2421845 and Simons Foundation MPS-AI-00010513.

This research used the DeltaAI advanced computing and data resource, which is supported by the National Science Foundation (award OAC 2320345) and the State of Illinois. DeltaAI is a joint effort of the University of Illinois Urbana-Champaign and its National Center for Supercomputing Applications. 

This research was supported in part through the computational resources and staff contributions provided for the Quest high performance computing facility at Northwestern University, which is jointly supported by the Office of the Provost, the Office for Research, and Northwestern University Information Technology.

This research was supported in part by the Illinois Computes project which is supported by the University of Illinois Urbana-Champaign.

We are grateful to Nabeel Rehemtulla for valuable discussions. We thank the anonymous reviewers for their constructive feedback.

% Which model performs the best?
% Is this sufficient for use with LSST Alerts?
% How can this model generalize to other time-series

\bibliographystyle{plainnat}
\bibliography{aaai2026}

\end{document}